\documentclass[accepted]{uai2023} 

\usepackage[american]{babel}

\usepackage{natbib} 
\usepackage{mathtools} 
\usepackage{booktabs} 
\usepackage{caption}
\usepackage{subcaption}
\usepackage{tikz} 
\usetikzlibrary{automata,arrows,positioning,calc}
\usetikzlibrary{shapes.multipart}
\usetikzlibrary{snakes}
\newcommand{\mysecref}[1]{\S\ref{#1}}
\usepackage{pifont}
\newcommand{\cmark}{\ding{51}}%
\newcommand{\xmark}{\ding{55}}%
\usepackage[ruled,vlined]{algorithm2e}

\usepackage{amsmath,amsfonts,bm}

\def\eqref#1{equation~\ref{#1}}

\def\1{\bm{1}}

\DeclareMathAlphabet{\mathsfit}{\encodingdefault}{\sfdefault}{m}{sl}
\SetMathAlphabet{\mathsfit}{bold}{\encodingdefault}{\sfdefault}{bx}{n}

\def\gV{{\mathcal{V}}}

\def\sR{{\mathbb{R}}}

\newcommand{\E}{\mathbb{E}}

\usepackage{float} 
\usepackage{multirow}
\usepackage[capitalize,noabbrev]{cleveref}

\usepackage{xcolor}

    \makeatletter
\def\@fnsymbol#1{\ensuremath{\ifcase#1\or \dagger\or \ddagger\or
   \mathsection\or \mathparagraph\or \|\or **\or \dagger\dagger
   \or \ddagger\ddagger \else\@ctrerr\fi}}
    \makeatother
\title{Molecule Design by Latent Space Energy-Based Modeling and Gradual Distribution Shifting}

\author[1]{Deqian Kong\thanks{Equal contribution}}
\author[2]{Bo Pang$^\dagger$}
\author[3]{Tian Han}
\author[1]{Ying Nian Wu}
\affil[1]{%
    Department of Statistics\\
    University of California, Los Angeles
}
\affil[2]{%
    Salesforce Research
}
\affil[3]{%
    Department of Computer Science\\ 
    Stevens Institute of Technology
  }

\begin{document}
\maketitle

\begin{abstract}
  Generation of molecules with desired chemical and biological properties such as high drug-likeness, high binding affinity to target proteins, is critical for drug discovery. In this paper, we propose a probabilistic generative model to capture the joint distribution of molecules and their properties. Our model assumes an energy-based model (EBM) in the latent space. Conditional on the latent vector, the molecule and its properties are modeled by a molecule generation model and a property regression model respectively.  To search for molecules with desired properties,  we propose a sampling with gradual distribution shifting (SGDS) algorithm, so that after learning the model initially on the training data of existing molecules and their properties, the proposed algorithm gradually shifts the model distribution towards the region supported by molecules with desired values of properties. Our experiments show that our method achieves very strong performances on various molecule design tasks. The code and checkpoints are available at~\url{https://github.com/deqiankong/SGDS}. 
\end{abstract}

\section{Introduction}

In drug discovery, it is of vital importance to find or design molecules with desired pharmacologic or chemical properties such as high drug-likeness and binding affinity to a target protein. It is challenging to directly optimize or search over the drug-like molecule space since it is discrete and enormous, with an estimated size on the order of $10^{33}$~\citep{polishchuk2013estimation}. 

Recently, a large body of work attempts to tackle this problem. The first line of work leverages deep generative models to map the discrete molecule space to a continuous latent space, and optimizes molecular properties in the latent space with methods such as Bayesian optimization \citep{gomez2018automatic, kusner2017grammar, jin2018junction}. The second line of work recruits reinforcement learning algorithms to optimize properties in the molecular graph space directly \citep{you2018graph, de2018molgan, zhou2019optimization, shi2020graphaf, luo2021graphdf}. A number of other methods have been proposed to optimize molecular properties with genetic algorithms~\citep{nigam2020augmenting}, particle-swarm algorithms~\citep{winter2019efficient}, and specialized MCMC methods \citep{xie2021mars}. 

In this work, we propose a method along the first line mentioned above, by learning a probabilistic latent space generative model of molecules and optimizing molecular properties in the latent space. Given the central role of latent variables in this approach, we emphasize that it is critical to learn a latent space model that captures the data regularities of the molecules. Thus, instead of assuming a simple Gaussian distribution in the latent space as in prior work~\citep{gomez2018automatic, jin2018junction}, we assume a flexible and expressive energy-based model (EBM)~\citep{lecun2006tutorial,ngiam2011learning,kim2016deep,xie2016theory,kumar2019maximum,nijkamp2019learning,du2019implicit,grathwohl2019your,finn2016connection} in latent space. This leads to a \textit{latent space energy-based model} (LSEBM) as studied in \cite{pang2020learning,nie2021controllable}, where LSEBM has been shown to model the distributions of natural images and text well. 

\begin{figure*}[h]
\begin{subfigure}[t]{.3\linewidth}
    \centering
	\begin{tikzpicture}[->, >=stealth', auto, thick, node distance=1.4cm]
	\tikzstyle{every state}=[fill=white,draw=black,thick,text=black,scale=1,minimum size=0.75cm]
	\node[state]    (x)[fill=black!20]                  {$x$};
	\node[state]    (z)[above of=x]                     {$z$};
	\node[state]    (y)[above of=z,fill=black!20]       {$y$};
	\path
	(z) edge[left]      node{\small $p_\gamma(y|z)$}         (y)
	(z) edge[left]      node{\small $p_\beta(x|z)$}          (x)
	(y) edge[bend left, dotted]      node[align=center]{\scriptsize\texttt{Conditional}\\\scriptsize\texttt{Generation}}          (z);
	\node       (pa)[right= 0.1cm of z]    {\small $p_\alpha(z)$};
	\end{tikzpicture}
	\caption{Single-Objective Optimization.}
	\label{fig:SGDS2}
\end{subfigure}%
{\hskip 0.02in}
\begin{subfigure}[t]{.3\linewidth}
    \centering
	\begin{tikzpicture}[->, >=stealth', auto, thick, node distance=1.4cm]
	\tikzstyle{every state}=[fill=white,draw=black,thick,text=black,scale=1,minimum size=0.75cm]
	\node[state]    (x)[fill=black!20]                  {$x$};
	\node[state]    (z)[above = 0.5cm of x]                     {$z$};
    \node           (y)[above of=z]    {$\cdots$};
	\node[state]    (y1)[left=0.2cm of y,fill=black!20,label=center:$y_1$] {\phantom{$y$}};
	\node[state]    (ym)[right =0.2cm of y,fill=black!20,label=center:$y_m$] {\phantom{$y$}};
	\path
	(z) edge[left]      node{\scriptsize$p_\beta(x|z)$}          (x)
	(z) edge[right]      node[right]{\scriptsize $p_{\gamma_m}(y_m|z)$}                        (ym)
	(z) edge[left]      node{\scriptsize $p_{\gamma_1}(y_1|z)$}                        (y1);
	\node       (pa)[right= 0.1cm of z]    {\scriptsize$p_\alpha(z)$};
	\end{tikzpicture}
	\caption{Multi-Objective Optimization.}
	\label{fig:SGDS3}
\end{subfigure}%
{\hskip 0.02in}
\begin{subfigure}[t]{.4\linewidth}
    \centering
  \includegraphics[width=.7\textwidth]{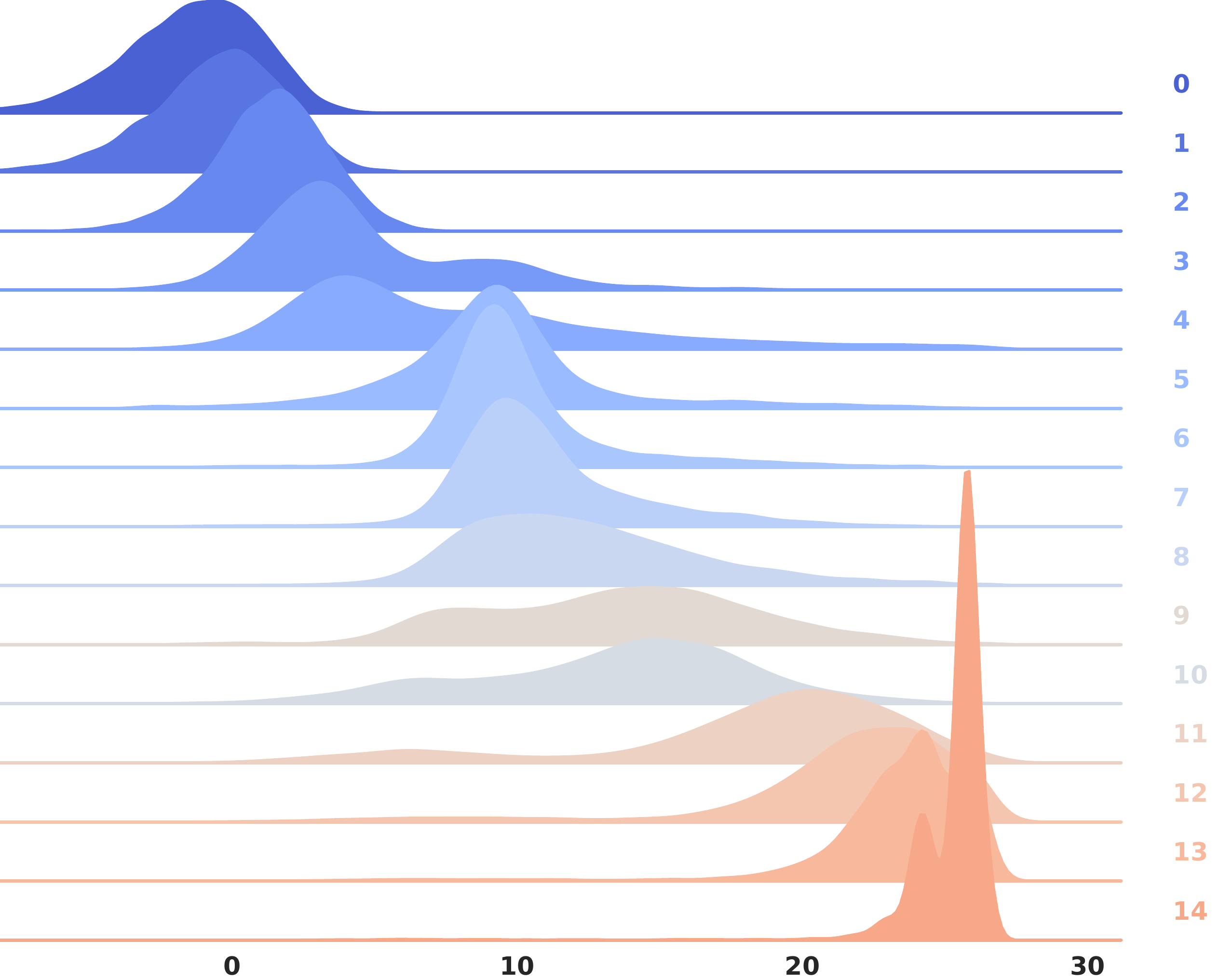}
  \caption{Shift of distribution of $y$.}
	\label{fig:SGDS4}
  \end{subfigure}%
\caption{An illustration of the joint distribution of molecule with its single property (a) or multiple properties (b). $x$ represents a molecule, $z$ is the latent vector, $y$ is a molecular property of interest, $\{y_j\}_{j=1}^m$ indicates $m$ properties. (c) illustrates the shift of the distribution of a single property in sampling with gradual distribution shifting (SGDS).}
\label{fig:SGDS}
\end{figure*}
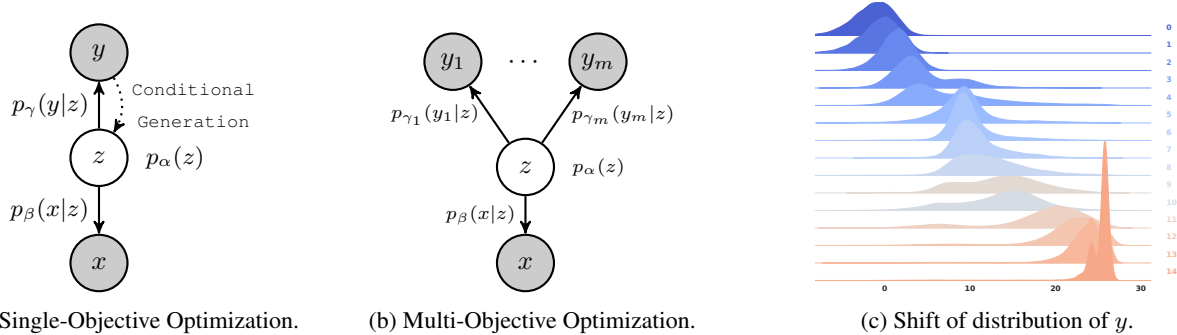

Going beyond existing latent space energy-based models  \cite{pang2020learning,nie2021controllable}, our work makes two innovations:

First, given our goal of property optimization, we learn a joint distribution of molecules and their properties. Our model consists of (1) an energy-based model (EBM) in a low-dimensional continuous latent space, (2) a molecule generation model that generates molecule given the latent vector, and (3) a property regression model that predicts the value of the property given the latent vector. See \cref{fig:SGDS2} for an illustration of the model.   We first learn the initial model on the training data that consist of existing molecules and their properties. All three components in our model are learned jointly by an approximate maximum likelihood algorithm.

Second, and more importantly, we propose a {\em sampling with gradual distribution shifting} (SGDS) method for molecule design. We first sample molecules and their property values from the initial model learned on the training data mentioned above. Then we gradually shift the joint distribution  towards the region supported by molecules with high property values. Specifically, our method iterates the following steps. (1) Shift the sampled property values by a small constant towards the desired target value. (2) Generate molecules given the shifted property values. (3) Obtain the ground-truth property values of the generated molecules by querying the software. (4) Update the model parameters by learning from the generated molecules and their ground-truth property values.  Because of the flexibility of the latent space energy-based model, the model can be updated to account for the change of the joint distribution of the generated molecules and their ground-truth property values in the gradual shifting process. \cref{fig:SGDS4} illustrates the shifting of the distribution of the property values of the generated molecules. 

In drug discovery, most often we need to consider multiple properties simultaneously. Our model can be extended to this setting straightforwardly. With our method, we only need to add a regression model for each property, while the learning and sampling methods remain the same (see ~\cref{fig:SGDS4}). We can then simultaneously shift the values of the multiple properties for multi-objective optimization. 

We evaluate our method in various settings including single-objective and multi-objective optimization. Our method outperforms prior methods by significant margins. 

In summary, our contributions are as follows: 
\begin{itemize}[leftmargin=*]
    \itemsep0em 
    \item We propose to learn a latent space energy-based model for the joint distribution of molecules and their properties.
    \item We develop a {sampling with gradual distribution shifting} method, which enables us to extrapolate the data distribution and sample from the region supported by molecules with high property values. 
    \item Our methods are versatile enough to be extended to optimizing multiple properties simultaneously.
    \item Our model achieves state-of-the-art performances on a range of molecule optimization tasks.
\end{itemize}

{\bf Caveat.} As in most existing work on molecule design, we assume that the value of a property of interest of a given molecule can be obtained by querying an existing software. There are two research problems in this endeavor. (1) Developing software that can output biologically or chemically accurate value of the property for an input molecule. (2) Developing method that can optimize the property values output by a given software. While problem (1) is critically important, our work is exclusively about problem (2). We duly acknowledge that existing software may need much improvements. Meanwhile our method can be readily applied to the improved versions of software. 

\section{Related Work}

\textbf{Optimization with Generative Models.}
Deep generative models approximate the distribution of molecules with desired biological or non-biological properties. Existing approaches for generating molecules include applying variational autoencoder (VAE)~\citep{kingma2013auto} and generative adversarial network (GAN) \citep{goodfellow2014generative} etc. to molecule data~\citep{gomez2018automatic,jin2018junction,de2018molgan,honda2019graph,madhawa2019graphnvp,shi2020graphaf,zang2020moflow,kotsias2020direct,chen2021molecule,fu2020core,liu2021graphebm,bagal2021liggpt,eckmann2022limo,segler2018generating}. After learning continuous representations for molecules, they are further able to optimize using different methods. \citep{segler2018generating} proposes to optimize by simulating design-synthesis-test cycles. \citep{gomez2018automatic, jin2018junction,eckmann2022limo} propose to learn a surrogate function to predict properties, and then use Bayesian optimization to optimize the latent vectors. However, the performance of this latent optimization is not satisfactory due to three major issues. First, it is difficult to train an accurate surrogate predictor especially for those novel molecules with high properties along the design trajectories. Second, as the learned latent space tries to cover the fixed data space, its ability to explore the targets out of the distribution is limited~\citep{guacamol,huang2021therapeutics}. Third, those methods are heavily dependent on the quality of learned latent space, which requires non-trivial efforts to design encoders when dealing with multiple properties. To address the above issues, \citep{eckmann2022limo} use VAE to learn the latent space and train predictors separately using generated molecules, and then leverage latent inceptionism, which involves the decoder solely, to optimize the latent vector with multiple predictors. In this paper, we propose an encoder-free model in both training and optimization to learn the joint distribution of molecules and properties. We then design an efficient algorithm to shift the learned distribution gradually. 

\textbf{Optimization with Reinforcement Learning and Evolutionary Algorithms.}
Reinforcement learning (RL) based methods directly optimize and generate molecules in an explicit data space~\citep{you2018graph,zhou2019optimization,jin2020multi,gottipati2020learning}. By formulating the property design as a discrete optimization task, they can modify the molecular substructures guided by an oracle reward function. However, the training of those RL-based methods can be viewed as rejection sampling which is difficult and inefficient due to the random-walk search behavior in the discrete space. Evolutionary algorithms (EA) also formulate the optimization in a discrete manner~\citep{nigam2020augmenting,jensen2019graph,xie2021mars,fu2021differentiable,fu2021mimosa}. By leveraging carefully-crafted combinatorial search algorithms, they can search the molecule graph space in a flexible and efficient way. However, the design of those algorithms is non-trivial and domain specific.

\section{Methods}
\subsection{Problem Setup and Overview}
We use the SELFIES representation for molecules~\citep{krenn2020self}. It encodes each molecule as a string of characters and ensures validity of all SELFIES strings. Let $x = (x^{(1)}, ..., x^{(t)},..., x^{(T)})$ be a molecule string encoded in SELFIES, where $x^{(t)} \in \gV$ is the $t$-th character and $\gV$ is the vocabulary.

Suppose $y \in \sR$ represents a molecular property of interest. 
Then the problem we attempt to tackle is to optimize $x$ such that its property $y = y^*$ where $y^*$ is some desirable value for $y$. We take a probabilistic approach and treat the optimization problem as a sampling problem, that is, 
\begin{equation}
\label{eq:t1}
 x^* \sim p(x | y = y^*).   
\end{equation}
This is a \textit{single-objective optimization} problem since only one property is targeted. In real-world drug design settings, we are more likely to optimize multiple properties simultaneously, that is, \textit{multi-objective optimization}. Suppose we optimize for $\{y_j \in \sR \}_{j=1}^m$, then our task is to sample
\begin{equation}
\label{eq:t2}
 x^* \sim p(x | y_1 = y_1^*, ..., y_m = y_m^*).   
\end{equation}

To address these problems, we propose a solution within a unified probabilistic framework. As a first step, we need to model or approximate the data distribution of molecules and their properties, $p_{\rm data}(x, y)$. 
To this end, we recruit latent space energy-based model (LSEBM) \citep{pang2020learning,nie2021controllable} to model the molecule and properties. LSEBM assumes that a latent vector $z \in \sR^d$ in a low dimensional latent space follows an energy-based prior model $p(z)$. Conditional on $z$, the molecule $x$ and the property $y$ are independent, so that the joint distribution $p(x, y, z)$ can be factorized as $p(z)p(x|z)p(y|z)$, which leads to $p(x, y) = \int p(z) p(x|z) p(y|z) dz$ as an approximation to $p_{\rm data}(x, y)$. 
The latent space energy-based prior model $p(z)$, the molecule generation model $p(x|z)$, and the property regression model $p(y|z)$ can be jointly learned by an  approximate maximum likelihood algorithm (see \mysecref{sec: learning} and ~\cref{algo:learning}). LSEBM within the context of molecule data is presented in \mysecref{sec:soo}. 

For the purpose of property optimization, we are required to generate molecules $x$ with some desirable property $y^*$. Rather than direct optimization in the molecule space, we choose to optimize $z$ in the latent space. We first consider the single-objective optimization problem (\cref{eq:t1}). 
 With the learned model, we propose to optimize $x$ given $y=y^*$ by ancestral sampling,
\begin{align}
    z^* \sim p(z | y = y^*),\quad x^* \sim p(x | z=z^*).
\end{align}
 However, if $y^*$ deviates from the observed data distribution of $y$, this naive solution involves sampling in an extrapolated regime (or out of distribution regime) where $y^*$ is not in the effective support of the learned distribution. To address this problem, we propose a \textit{Sampling with Gradual Distribution Shifting} (SGDS) approach where we gradually shift the learned distribution to a region where it is supported by high property values (see \mysecref{sec:sgds} and Algorithm \ref{algo:sgds}). 

Our model is designed to be versatile such that it admits straightforward extension to multi-objective optimization. To optimize $x$ given $\{y_j = y_j^*\}_{j=1}^m$, we can simply augment the joint distribution with more regression models, i.e., $p(x, z, y_1, ..., y_m) = p(z) p(x|z) \prod_{j=1}^m p(y_j|z)$. The optimization procedure follows the same SGDS approach. See \mysecref{sec:moo} for more details on multi-objective optimization.

\subsection{Joint Distribution of Molecule and Molecular Property}
\label{sec:soo}
Suppose $x = (x^{(1)}, ..., x^{(t)},..., x^{(T)})$ is a molecule string in SELFIES, $y \in \mathbb{R}$ is the target property of interest, and $z \in \mathbb{R}^d$ is the latent vector. Consider the following model, 
\begin{align}
z \sim p_{\alpha}(z), \quad [x \mid z] \sim p_\beta(x|z), \quad [y \mid z] \sim p_\gamma(y|z),
\end{align}
where $p_{\alpha}(z)$ is a prior model with parameters $\alpha$, $p_\beta(x|z)$ is a molecule generation model with parameters $\beta$, and $p_\gamma(y|z)$ is a property regression model with parameter $\gamma$. In VAE~\citep{kingma2013auto}, the prior is simply assumed to be an isotropic Gaussian distribution. In our model, $p_{\alpha}(z)$ is formulated as a learnable energy-based model,
\begin{align}
p_{\alpha}(z) = \frac{1}{Z(\alpha)} \exp(f_{\alpha}(z)) p_0(z), \label{eq:prior}
\end{align}   
where $p_0(z)$ is a reference distribution, assumed to be isotropic Gaussian as in VAE. $f_{\alpha}: \sR^{d} \rightarrow \sR$ is a scalar-valued negative energy function and is parameterized by a small multi-layer perceptron (MLP) with parameters $\alpha$. $Z(\alpha) = \int \exp(f_\alpha (z)) p_0(z) dz = \E_{p_0}[\exp(f_\alpha(z))]$ is the normalizing constant or partition function. 

The molecule generation model $p_{\beta}(x|z)$ is a conditional autoregressive model,
\begin{align} 
    p_\beta(x|z) = \prod_{t=1}^T p_\beta(x^{(t)}|x^{(1)}, ..., x^{(t-1)}, z) 
\end{align} 
which is parameterized by a one-layer LSTM \cite{hochreiter1997long} with parameters $\beta$. Note that the latent vector $z$ controls every step of the autoregressive model. It is worth pointing out the simplicity of the molecule generation model of our method considering that those in prior work involve complicated graph search algorithm or alternating generation of atoms and bonds with multiple networks. 

Given a molecule $x$, suppose $y$ is the chemical property of interest, such as drug likeliness or protein  binding affinity. The ground-truth property value can be computed for an input $x$ via open-sourced software such as RDKit~\citep{landrum2013rdkit} and AutoDock-GPU~\citep{santos2021accelerating}. We assume that given $z$, $x$ and $y$ are conditionally independent, so that 
\begin{align}
p_{\theta}(x, y, z) &= p_{\alpha}(z) p_\beta(x|z) p_\gamma( y|z),
\label{eq:joint}
\end{align}
where $\theta = (\alpha, \beta, \gamma)$. We use the model $p_\theta(x, y) = \int p_\theta(x, y, z)dz $ to approximate the data distribution  $p_\mathrm{data}(x, y)$. See Supplement for a detailed discussion. 

The property regression model can be written as
\begin{align}
   p_{\gamma}(y|z) = \frac{1}{\sqrt{2 \pi \sigma^2}} \exp\left( - \frac{1}{2\sigma^2} (y - s_\gamma(z) )^2\right), 
\end{align}
where $s_\gamma(z)$ is a small MLP, with parameters $\gamma$, predicting $y$ based on the latent $z$. The variance $\sigma^2$ is set as a constant or a hyperparameter in our work. 

\subsection{Learning Joint Distribution}
\label{sec: learning}

Suppose we observe training examples $\{(x_i, y_i), i = 1, ..., n\}$. The log-likelihood function is $L( \theta) = \sum_{i=1}^{n} \log p_\theta(x_i, y_i)$.
The learning gradient can be calculated according to 
\begin{align} 
   &\nabla_\theta  \log p_\theta(x, y)  = \E_{p_\theta(z|x, y)} \left[ \nabla_\theta \log p_\theta(x, y, z) \right] \nonumber\\
     &= \E_{p_\theta(z|x, y)} \left[ \nabla_\theta (\log p_\alpha(z) + \log p_\beta(x|z) + \log p_\gamma(y|z)) \right].
\end{align}

For the prior model, 
\begin{align}
    \nabla_\alpha \log p_\alpha(z)  =  \nabla_\alpha f_\alpha(z) - \E_{p_\alpha(z)}[ \nabla_\alpha f_\alpha(z)]. 
\end{align}
The learning gradient given an example $(x, y)$ is
\begin{align} 
  &\delta_\alpha(x, y) =   \nabla_\alpha \log p_\theta(x, y) \nonumber\\&= \E_{p_\theta(z|x, y)}[\nabla_\alpha f_\alpha(z)] - \E_{p_\alpha(z)} [\nabla_\alpha f_\alpha(z)]. \label{eq:alpha}
\end{align}
Thus $\alpha$ is updated based on the difference between $z$ inferred from empirical observation $(x, y)$, and $z$ sampled from the current prior. 

For the molecule generation model,
\begin{align} 
\label{eq:beta}
\begin{split}
  \delta_\beta(x, y) =  \nabla_\beta \log p_\theta(x, y)  = \E_{p_\theta(z|x, y)} [\nabla_\beta \log p_{\beta}(x|z)].
\end{split}
\end{align} 
Similarly, for the property regression model, 
\begin{align} 
\label{eq:gamma}
\begin{split}
  \delta_\gamma(x, y) =  \nabla_\gamma \log p_\theta(x, y)  = \E_{p_\theta(z|x, y)} [\nabla_\gamma \log p_{\gamma}(y|z)].
\end{split}
\end{align} 

Estimating expectations in Equations \ref{eq:alpha}, \ref{eq:beta}, and \ref{eq:gamma} requires MCMC sampling of the prior model $p_\alpha(z)$ and the posterior distribution $p_\theta(z|x,y)$. We recruit Langevin dynamics \citep{neal2011mcmc,han2017abp}. For a target distribution $\pi(z)$, the dynamics iterates
\begin{align} 
z_{\tau+1} = z_\tau + s \nabla_z \log \pi(z_\tau) + \sqrt{2s} \epsilon_\tau, 
 \label{eq:Langevin}
\end{align}
where $\tau$ indexes the time step of the Langevin dynamics, $s$ is step size, and $\epsilon_\tau \sim {\mathcal N}(0, I_d)$ is the Gaussian white noise. $\pi(z)$ can be either the prior $p_\alpha(z)$ or the posterior $p_\theta(z|x, y)$. In either case, $\nabla_z \log \pi(z)$ can be efficiently computed by back-propagation. 

We initialize $z_0 \sim {\mathcal N}(0, I_d)$, and we run $\Gamma$ steps of Langevin dynamics (e.g. $\Gamma=20$) to approximately sample from the prior and the posterior distributions. The resulting learning algorithm is an approximate maximum likelihood learning algorithm. See \citep{pang2020learning,nijkamp2020learning2} for a theoretical understanding of the learning algorithm based on the finite-step MCMC. See also \citep{gao2020learning,yu2022latent} for learning EBMs at multiple noise levels for effective modeling and sampling of multimodal density. 

The learning algorithm is summarized in \cref{algo:learning}.

\begin{algorithm}
	\SetKwInOut{Input}{input} \SetKwInOut{Output}{output}
	\DontPrintSemicolon
	\Input{Learning iterations~$T$, learning rates for the prior, generation, and regression models $\{\eta_0, \eta_1, \eta_2\}$, initial parameters~$\theta_0 = (\alpha_0, \beta_0, \gamma_0)$, observed examples~$\{(x_i, y_i )\}_{i=1}^n$, batch size~$m$, number of prior and posterior sampling steps $\{\Gamma_0, \Gamma_1\}$, and prior and posterior sampling step sizes $\{s_0, s_1\}$.}
	\Output{ $\theta_T = (\alpha_{T}, \beta_{T}, \gamma_{T})$.}
	\For{$t = 0:T-1$}{			
		\smallskip
		1. {\bf Mini-batch}: Sample observed examples $\{ (x_i, y_i) \}_{i=1}^m$. \\
		2. {\bf Prior sampling}: For each $i$, sample $z_i^{-} \sim {p}_{\alpha_t}(z)$ using \cref{eq:Langevin}, where the target distribution $\pi(z) = {p}_{\alpha_t}(z)$, and $s = s_0$, $\Gamma = \Gamma_0$. \\
		3. {\bf Posterior sampling}: For each $(x_i, y_i)$, sample $z_i^{+} \sim {p}_{\theta_t}(z|x_i, y_i)$ using~\cref{eq:Langevin}, where the target distribution $\pi(z) = {p}_{\theta_t}(z|x_i, y_i)$, and $s = s_1$, $\Gamma = \Gamma_1$. \\
		4. {\bf Update prior model}: $\alpha_{t+1} = \alpha_t + \eta_0 \frac{1}{m} \sum_{i=1}^{m} [\nabla_\alpha f_{\alpha_t}(z_i^{+}) - \nabla_\alpha f_{\alpha_t}(z_i^{-})]$. \\
		5. {\bf Update generation model}: $\beta_{t+1} = \beta_t + \eta_1 \frac{1}{m} \sum_{i=1}^{m}\nabla_\beta \log p_{\beta_t}(x_i | z_i^{+})$. \\ 
		6. {\bf Update regression model}: $\gamma_{t+1} = \gamma_t + \eta_2 \frac{1}{m} \sum_{i=1}^{m}\nabla_\gamma \log p_{\gamma_t}(y_i | z_i^{+})$. 
		}
	\caption{Learning joint distribution.}
	\label{algo:learning}
\end{algorithm}

\subsection{Sampling with Gradual Distribution Shifting (SGDS)}
\label{sec:sgds}
To tackle the single-objective optimization problem (\cref{eq:t1}), one naive approach is to perform ancestral sampling with two steps, given some desirable property value $y^*$,
\begin{align}
\label{eq:naive}
    &(\mathrm{i}) \  z^* \sim p_\theta(z|y=y^*) \propto p_\alpha(z) p_\gamma(y=y^*|z), \\ &(\mathrm{ii}) \ x^* \sim p_\beta(x|z=z^*), 
\end{align} 
where $(\mathrm{i})$ is an application of Bayes rule, with $p_\alpha(z)$ as the prior and $p_\gamma(y|z)$ as the likelihood. Sampling from $p_\theta(z|y)$ can be carried out by Langevin dynamics in~\cref{eq:Langevin} by replacing the target distribution $\pi(z)$ with $p_\theta(z|y)$.

Our model $p_\theta(x, y, z)$ is learned to capture the data distribution. In real-world settings, $y^*$ might not be within the support of the data distribution. Therefore, sampling following~\cref{eq:naive} does not work well since it involves extrapolating the learned distribution. 

We propose an iterative updating method called \textit{sampling with gradual distribution shifting} (SGDS) to address this issue. In particular, we first leverage the $n$ samples collected from the common dataset  $\{(x_i^{0}, y_i^{0})\}_{i=1}^n$ (e.g. ZINC, $n = 250,000$) to learn the initial joint distribution $p_{\theta_0}(x,y)$ as a valid starting point. Then we shift the joint distribution progressively using a smaller number $k$ (e.g., $k = 10,000$) of synthesized samples $\{(x_i^{t}, y_i^{t})\}_{i=1}^{k}$ from distribution $p_{\theta_{t-1}}$ at the previous iteration, where $k \ll n$. Therefore, by progressively learning the joint distribution with $T$ (e.g., $T = 30$) shift iterations, $p_{\theta_{1}},\dots,p_{\theta_{T}}$, at the last several iterations, we expect to generate molecules with desirable properties that are significantly distant from the initial distribution as shown in~\cref{fig:SGDS4}. 

Next, we shall explain in detail the steps to generate $\{(x_i^{t}, y_i^{t})\}_{i=1}^{k}$ given $p_{\theta_{t-1}}$. In a property maximization task, we shift the support slightly by adding a small $\Delta_y$ to all $y$'s, 
\begin{align}
    \tilde{y}^t = y^{t-1} + \Delta_y,
\end{align}
and generate $x^t$ conditional on shifted $\tilde{y}^t$, following \cref{eq:naive},
\begin{align}
\label{eq:sgds}
    (\mathrm{i}) \  &z^t \sim p_{\theta_{t-1}}(z|y=\tilde{y}^t), \\ (\mathrm{ii}) \ &x^t \sim p_{\beta_{t-1}}(x|z=z^t).
\end{align} 
$\Delta_y$ can be chosen as a fixed small value. After generating $x^t$, its ground-truth property value $y^t$ can be computed by calling the corresponding engines such as RDKit and AutoDock-GPU. In \cref{eq:sgds}, sampling can be achieved by langevin dynamics as in~\cref{eq:Langevin}. For the sake of efficiency, we propose to run persistent chain by initializing the Langevin dynamics from the latent vectors generated in the previous iteration \cite{han2017abp}. This is also called warm start. Specifically, we have 
\begin{align} 
&z_0^t = z^{t-1}_\Gamma,\nonumber \\
&z^t_{\tau+1} = z^t_\tau + s \nabla_z \log p_{\theta_{t-1}}(z|\tilde{y}^t) + \sqrt{2s} \epsilon_\tau,
\label{eq:warm_start}
\end{align}
for $\tau = 1, ..., \Gamma$, where $\Gamma$ is the length of Markov chain in each iteration. With warm start, we use $\Gamma=2$ in our experiments. 

For more efficient optimization via distribution shifting, we further introduce a rank-and-select scheme by maintaining a buffer of top-$k$ samples of $(z, x, y)$ (where $z$ is the sampled latent vector, $x$ is the generated molecule, and $y$ is the ground-truth property value of $x$). Specifically, we maintain a buffer which consists of $k$ samples of $(z, x, y)$ with the highest values of $y$ in the past shifting iterations. In each shift iteration, conditioned on the shifted property values, with warm start, initialized from these $k$ vectors $z$ in the buffer, a new  batch of $k$ latent vectors $z$, molecules $x$, and their ground-truth values $y$ can be produced in the new shift iteration. We rank all the $2k$ samples of $(z, x, y)$ (including $k$ newly generated ones and $k$ samples in the buffer) and select the top-$k$ samples of $(z, x, y)$ based on the ground-truth values $y$. The $k$ samples of $(z, x, y)$ in the buffer are then updated by those newly selected $k$ samples of $(z, x, y)$. We call this procedure as \textit{rank-and-select}. This rank-and-select procedure can also be applied to constrained optimization tasks, where we select those sampled molecules that satisfy the given constraints. With the selected samples, we then shift the model distribution by learning from these samples with several learning iterations. The SGDS algorithm is summarized in Algorithm~\ref{algo:sgds}.

\begin{algorithm}
	\SetKwInOut{Input}{input} \SetKwInOut{Output}{output}
	\DontPrintSemicolon
	\Input{Shift iterations~$T$, initial pretrained parameters $\theta_0 = (\alpha_0, \beta_0, \gamma_0)$, initial examples~$\{(x_i^0, y_i^0) \}_{i=1}^k$ from the data distribution boundary, shift magnitude $\Delta_y$, ${\rm PropertyComputeEngine} = {\rm RDKit}$ or ${\rm AutoDock}$-${\rm GPU}$, ${\rm LearningAlgorithm} = {\rm Algorithm}$~\ref{algo:learning}.}
	\Output{ $\{(x_i^T, y_i^T)\}_{i=1}^k$.}
	\For{$t = 1:T$}{			
		\smallskip
		1. {\bf Property shift}: For each $y_i^{t-1}$, $\tilde{y}_i^{t} = y_i^{t-1} + \Delta_y$. \\
		2. {\bf Latent sampling with warm start}: For each $\tilde{y}_i^{t}$, sample $z_i^{t} \sim p_{\theta_{t-1}}(z | \tilde{y}_i^{t})$ using~\cref{eq:warm_start}. \\ 
		3. {\bf Molecule generation}: For each $z_i^{t}$, sample $x_i^{t} \sim p_{\theta_{t-1}}(x|z_i^{t})$. \\ 
		4. {\bf Property computation}: For each $x_i^{t}$, compute $y_i^{t} = {\rm PropertyComputeEngine} (x_i^{t})$. \\
        5. {\bf Rank-and-select:} Update the buffer of top-$k$ samples $\{z_i^{t}, x_i^{t}, y_i^{t}\}_{i=1}^{k}$ by rank-and-select. \\
		6. {\bf Distribution shift}: $\theta_{t} = {\rm LearningAlgorithm} (\{ (x_i^{t}, y_i^{t}) \}_{i=1}^k, \theta_{t-1}) $. 
		}
	\caption{SGDS for single property optimization.}
	\label{algo:sgds}
\end{algorithm}

\subsection{Multi-Objective Optimization}
\label{sec:moo}

We next consider the multi-objective optimization problem. Suppose we optimize for a set of properties $\{y_j\}_{j=1}^m$, then we learn a property regression model for each property $y_j$,
\begin{equation}
    p_{\gamma_j}(y_j | z) = \frac{1}{\sqrt{2 \pi \sigma_j^2}} \exp\left( - \frac{1}{2\sigma_j^2} (y_j - s_{\gamma_j}(z) )^2\right), 
\end{equation}
where each $s_{\gamma_j}$ is a small MLP with parameters $\gamma_j$.
We assume that given $z$ the properties are conditionally independent, so the joint distribution is
\begin{equation}
    p_\theta(x, z, y_1, ..., y_m) = p_\alpha(z) p_\beta(x | z) \prod_{j=1}^m p_{\gamma_j}(y_i | z).
\end{equation}

Under our framework, both the learning and the sampling algorithm for the single-objective problem can be straightforwardly extended to the multi-objective setting. In SGDS, we shift the values of the multiple properties simultaneously, and generate molecules conditional on the multiple properties. 

\section{Experiments}
To demonstrate the effectiveness of our proposed method, SGDS, we compare our model with previous SOTA methods for molecule design including single-objective optimization (\mysecref{sub:single_property_optimization}), multi-objective optimization (\mysecref{sub:multi_objective_binding_affinity_maximization}) and constrained optimization (\mysecref{sec:constrained}). In molecule design experiments, we consider both non-biological and biological properties. Finally, we add ablation studies to analyze the effects of different components in SGDS. We also conduct unconditional molecule generation experiments for sanity check of the model and discuss the mode traversing in the latent space at the end of this section.

\subsection{Experimental Setup}
\textbf{Datasets.} For the molecule property optimization task, we report results on ZINC~\citep{irwin2012zinc} and MOSES~\citep{polykovskiy2020molecular}, which consist of around $250$k and $2$M molecules respectively. 

Encoding systems in molecular studies typically include SMILES~\citep{weininger1988smiles}, SELFIES~\citep{krenn2020self}, and graph representations. SMILES and SELFIES linearize a molecular graph into character strings. SMILES has historically faced challenges regarding validity (the percentage of molecules that satisfy the chemical valency rules). Recently, SELFIES was introduced, offering an encoding system where each string inherently corresponds to a valid molecule. 
We use SELFIES representation in our work. 

The non-biological properties (such as penalized logP, QED, etc.) can be computed using RDKit~\citep{landrum2013rdkit}. Following~\citep{eckmann2022limo}, we use the docking scores from AutoDock-GPU~\citep{santos2021accelerating} to approximate the binding affinity to two protein targets, human estrogen receptor (ESR1) and human peroxisomal acetyl-CoA acyl transferase 1 (ACAA1).




\textbf{Training Details.} There are three modules in our method, the molecule generation model $p_\beta(x|z)$, the energy-based prior model $p_\alpha(z)$, and property regression model $\{p_{\gamma_j}(y|z)\}_{j=1}^m$, where $m$ is the total number of properties we aim to optimize. The generation model $p_\beta(x|z)$ is parameterized by a single-layer LSTM with $1024$ hidden units where the dimension of latent vector $z$ is $100$. The energy-based prior model $p_\alpha(z)$ is a $3$-layer MLP. Each of the property regression model $p_{\gamma_j}(y|z)$ is a 3-layer MLP. It is worth mentioning that compared to most previous models, SGDS is characterized by its simplicity without adding inference networks for sampling, or RL-related modules for optimization. In order to get valid initial distribution $\theta_0$ for SGDS, we first train our model for $30$ epochs on ZINC. We use Adam optimizer~\citep{kingma2015adam} to train our models with learning rates $10^{-4}$ for energy-based prior model, and $10^{-3}$ for the molecule generation model and property regression model. During SGDS, we use $30$ shifting iterations for single-objective optimization and $20$ for multi-objective optimization. For each iteration of distribution shifting, we sample $10^4$ boundary examples except binding affinity, where $2\times10^3$ examples are used to speed up calculation, and then we update the model parameters $\theta$ for $10$ iterations using \cref{algo:learning} with Adam optimizer and the same learning rates mentioned above. All experiments are conducted on Nvidia Titan XP GPU.

\subsection{Single-Objective Optimization} 
\label{sub:single_property_optimization}
\textbf{Penalized logP and QED Maximization.} For non-biological properties, we are interested in Penalized logP and QED, both of which can be calculated by RDKit~\citep{landrum2013rdkit}.

Since we know Penalized logP scores have a positive relationship with the lengths of molecules, we maximize Penalized logP either with or without maximum length limit. Following~\citep{eckmann2022limo}, the maximum length is set to be the maximum length of molecules in ZINC using SELFIES. From Table~\ref{tab:single_non_bio}, we can see that with length limit, SGDS outperforms previous methods by a large margin. We also achieve the highest QED with/without length limit. These observations demonstrate the effectiveness of our method. We also illustrate our distribution shifting method in \cref{fig:SGDS4}. One can notice that, the distribution of the property is gradually shifted towards the region with higher values, and the final distribution is significantly distant from the initial one.

\begin{table}[h]
\caption{Non-biological single-objective optimization. Report top-3 highest scores found by each model. LL (Length Limit) denotes whether it has the maximum length limit. Baseline results obtained from~\citep{eckmann2022limo,you2018graph,luo2021graphdf,xie2021mars}.}
\label{tab:single_non_bio}
\resizebox{\linewidth}{!}{%
\centering
\begin{tabular}{lccccccc}
\toprule
\multicolumn{1}{c}{\multirow{2}{*}{\bf Method}} & \multicolumn{1}{l}{\multirow{2}{*}{\bf LL}} & \multicolumn{3}{c}{\bf Penalized logP $(\uparrow)$}                                          & \multicolumn{3}{c}{\bf QED $(\uparrow)$}        \\
\multicolumn{1}{c}{}                        & \multicolumn{1}{l}{}                    & \multicolumn{1}{c}{1st} & \multicolumn{1}{c}{2rd} & \multicolumn{1}{c}{3rd} & \multicolumn{1}{c}{1st} & \multicolumn{1}{c}{2rd} & \multicolumn{1}{c}{3rd} \\
\midrule
JT-VAE       & \xmark & 5.30 & 4.93 & 4.49 & 0.925 & 0.911 & 0.910 \\
GCPN         & \cmark & 7.98 & 7.85 & 7.80 & \bf 0.948 & 0.947 & 0.946 \\
MolDQN       & \cmark & 11.8 & 11.8 & 11.8 & \bf 0.948 & 0.943 & 0.943 \\
MARS         & \xmark & 45.0 & 44.3 & 43.8 & \bf 0.948 & \bf 0.948 & \bf 0.948\\
GraphDF      & \xmark & 13.7 & 13.2 & 13.2 & \bf 0.948 & \bf 0.948 & \bf 0.948 \\
LIMO         & \cmark & 10.5 & 9.69 & 9.60 & 0.947 & 0.946 & 0.945 \\
\midrule
\textbf{SGDS}         & \cmark & \bf26.4 &\bf 25.8 &\bf 25.5 &\bf 0.948 &\bf 0.948 &\bf 0.948\\
\textbf{SGDS}         & \xmark &\bf 158.0 &\bf 157.8 &\bf 157.5 &\bf 0.948 &\bf 0.948 &\bf 0.948
\end{tabular}
}
\end{table}
\textbf{Biological Property Optimization.} ESR1 and ACAA1 are two human proteins. We aim to design ligands (molecules) that have the maximum binding affinities towards those target proteins. ESR1 is well-studied, which has many existing binders. However, we do not use any binder-related information in SGDS. Binding affinity is measured by the estimated dissociation constants $\mathrm{K_D}$, which can be approximated by docking scores from AutoDock-GPU~\citep{santos2021accelerating}. Large binding affinities corresponds to small $\mathrm{K_D}$. That is, we aim to minimize $\mathrm{K_D}$.
Table~\ref{tab:single_bio} shows that our model outperforms previous methods on both ESR1 and ACAA1 binding affinity maximization tasks by  large margins. Comparing to existing methods, much more molecules with high binding affinities can be directly sampled from the last several shifting iterations. See Supplement for more examples. Producing those ligands with high binding affinity plays a vital role in the early stage of drug discovery. 

\begin{table}[h]
\caption{Biological single-objective optimization. Report top-3 lowest $\mathrm{K_D}$ (in nanomoles/liter) found by each model. Baseline results obtained from~\citep{eckmann2022limo}.}
\label{tab:single_bio}
\centering
\resizebox{0.8\linewidth}{!}{%
\begin{tabular}{lcccccc}
\toprule
\multicolumn{1}{c}{\multirow{2}{*}{\bf Method}} & \multicolumn{3}{c}{\bf ESR1 $\mathrm{K_D}$ $(\downarrow)$} & \multicolumn{3}{c}{\bf ACAA1 $\mathrm{K_D}$ $(\downarrow)$}\\
\multicolumn{1}{c}{}                                           & 1st           & 2rd           & 3rd          & 1st           & 2rd           & 3rd                                                   \\ \midrule
GCPN                                                            & 6.4           & 6.6           & 8.5          & 75            & 83            & 84                                                    \\
MolDQN                                                          & 373           & 588           & 1062         & 240           & 337           & 608                                                   \\
MARS                                                            & 25            & 47            & 51           & 370           & 520           & 590                                                  \\
GraphDF                                                         & 17            & 64            & 69           & 163           & 203           & 236                                                  \\
LIMO                                                            & 0.72          & 0.89          & 1.4          & 37            & 37            & 41                                                    \\ \hline
\textbf{SGDS}                                                            & \bf 0.03              &\bf 0.03              &\bf  0.04            & \bf 0.11              & \bf 0.11              & \bf 0.12                                                      
\end{tabular}}
\end{table}

\subsection{Multi-objective Optimization} 
\label{sub:multi_objective_binding_affinity_maximization}

\paragraph{Multi-objective Binding Affinity Optimization.} We consider maximizing binding affinity, QED and minimizing synthetic accessibility score (SA) simultaneously. Following \cite{eckmann2022limo}, we exclude molecules with abnormal behaviors~\footnote{ In particular, we exclude molecules with QED$\uparrow$ smaller than 0.4, SA$\downarrow$ larger than 5.5, and too small (less than 5 atoms) or too large (more than 6 atoms) chemical rings.} to encourage the joint distribution shifts towards a desirable region in terms of pharmacologic and synthetic properties. Those heuristics can be conveniently added in our \textit{rank-and-select} step.
Table~\ref{tab:multi} shows our multi-objective results compared to LIMO and GCPN. From the results, we can see that SGDS is able to find the ligands with desired properties while keeping the pharmacologic structures. For ESR1, we have two existing binders on the market, Tamoxifen and Raloxifene. Our designed ligands have similar QED and SA, with very low $\mathrm{K_D}$.  Compared to existing methods, SGDS obtains better results in overall adjustments. For ACAA1, we do not have any existing binders. Compared with prior SOTA methods, our optimized ligands outperform those by a large margin in terms of $\mathrm{K_D}$. When comparing with single-objective optimization, we find that multi-objective optimization is more complicated, but it may be more useful in real world molecule design. While we still need domain expertise to determine the effectiveness of those ligands discovered by SGDS, we believe the ability of SGDS to generate many high quality molecules given multiple metrics is extremely useful in the early stage of drug discovery.

\begin{table}
\caption{Muli-objective optimization for both ESR1 and ACAA1. Report Top-2 average scores related to $\mathrm{K_D}$ (in nmol/L), QED and SA. Baseline results obtained from~\citep{eckmann2022limo}.}
\label{tab:multi}
\resizebox{1\linewidth}{!}{
\centering
\begin{tabular}{lcccccc}
\toprule
\multicolumn{1}{c}{\multirow{2}{*}{\bf Ligand}} & \multicolumn{3}{c}{\bf ESR1}  & \multicolumn{3}{c}{\bf ACAA1}                                     \\
\multicolumn{1}{c}{}                        &$\mathrm{K_D}$ $\downarrow$            &\bf QED $\uparrow$                      &\bf SA $\downarrow$  &$\mathrm{K_D}$ $\downarrow$            &\bf QED $\uparrow$                      &\bf SA $\downarrow$                    \\\midrule\multicolumn{1}{l}{Tamoxifen}               & 87                   & 0.45                     &   2.0      & $-$ & $-$ & $-$               \\
\multicolumn{1}{l}{Raloxifene}              & $7.9\times10^6$      & 0.32 & 2.4  & $-$ & $-$ & $-$\\ \midrule
\multicolumn{1}{l}{GCPN $\texttt{1}^\texttt{st}$}                & 810                  & 0.43                     & 4.2   & 8500                 &  0.69                     & 4.2                  \\
\multicolumn{1}{l}{GCPN $\texttt{2}^\texttt{nd}$}                & $27000$      & 0.80                     & 3.7   & 8500                 & 0.54                     & 4.3                  \\
\multicolumn{1}{l}{LIMO $\texttt{1}^\texttt{st}$}                & 4.6                  & 0.43                     & 4.8   & 28                   & 0.57                     & 5.5                  \\
\multicolumn{1}{l}{LIMO $\texttt{2}^\texttt{nd}$}                & 2.8                  & 0.64                     & 4.9   & 31                   & 0.44                     & 4.9                  \\\midrule
\multicolumn{1}{l}{\textbf{SGDS} $\texttt{1}^\texttt{st}$}                &\bf 0.36                 & 0.44                     & 3.99   & \bf 4.55             & 0.56                     &  \bf 4.07                 \\
\multicolumn{1}{l}{\textbf{SGDS} $\texttt{2}^\texttt{nd}$}                & 1.28                 &  0.44                     & 3.86     & 5.67             & 0.60                     & 4.58        
\end{tabular}}
\end{table}

\subsection{Constrained Optimization}
To optimize single-objective under some constrains, we use the original SGDS steps and in \textit{rank-and-select}, we only keep the molecules that satisfy the constraints.
\label{sec:constrained}
\paragraph{Similarity-constrained Penalized logP Maximization}
Following JT-VAE~\citep{jin2018junction}, this experiment aims to generate molecules with high penalized logP while being similar to the target molecules. Similarity is measured by Tanimoto similarity
between Morgan fingerprints with a cutoff value $\delta$. We compare our results with previous SOTA method in \cref{tab:sim}. The results show that SGDS tends to obtain better results with weak constraints (i.e. $\delta=0, 0.2$) with $100\%$ success rate, since different from optimized property, the constraints are added implicitly.

\begin{table}[h]
\caption{Similarity-constrained optimization results. LIMO results obtained from~\citep{eckmann2022limo,luo2021graphdf}.}
\label{tab:sim}
\centering
\resizebox{1\linewidth}{!}{
\begin{tabular}{lcccccc}
\toprule
\multicolumn{1}{c}{\multirow{2}{*}{$\delta$}} & \multicolumn{2}{c}{GraphDF}
&\multicolumn{2}{c}{LIMO} &\multicolumn{2}{c}{SGDS}\\                    
&\multicolumn{1}{c}{Improv.} & \multicolumn{1}{c}{$\%$ Succ.}
&\multicolumn{1}{c}{Improv.} & \multicolumn{1}{c}{$\%$ Succ.}
&\multicolumn{1}{c}{Improv.} & \multicolumn{1}{c}{$\%$ Succ.}\\
\midrule
0.0    & $5.9\pm 2.0$ & 100 & $10.1\pm 2.3$ & 100 &$\mathbf{19.1}\pm2.1$ & 100\\
0.2    & $5.6\pm 1.7$ & 100 & $5.8\pm 2.6$ & 99.0 &$\mathbf{7.4}\pm 1.9$ & 100\\
0.4    & $\mathbf{4.1}\pm 1.4$ & 100 & $3.6\pm 2.3$ & 93.7 &$3.8\pm 1.4$ & 97.5\\
0.6   & $1.7\pm 1.2$ & 93.0 & $1.8\pm 2.0$ & 85.5 &$\mathbf{2.6}\pm 2.0$ & 95.6\\
\end{tabular}
}
\end{table}

\paragraph{logP targeting}
In \cref{tab:targeting}, comparing to previous methods, SGDS is able to get competitive diversity scores with significantly better success rate in both ranges. That is because after SGDS, our model is shifted towards the region that is supported by molecules satisfying the logP constraints. Due to the flexibility of our EBM prior, SGDS achieves high diversity scores while keeping most of the sampled molecules within the logP range.
\begin{table}[h]
\caption{logP targeting to a certain range~\citep{eckmann2022limo,you2018graph,luo2021graphdf,xie2021mars}.}
\label{tab:targeting}
\centering
\resizebox{0.8\linewidth}{!}{
\begin{tabular}{lcccc}
\toprule
\multicolumn{1}{c}{\multirow{2}{*}{\bf Method}} & \multicolumn{2}{c}{$-2.5 \le$ logP $\le -2$}                                          & \multicolumn{2}{c}{$5 \le$ logP $\le 5.5$}        \\                         & \multicolumn{1}{c}{Success} & \multicolumn{1}{c}{Diversity} & \multicolumn{1}{c}{Success} & \multicolumn{1}{c}{Diversity}\\
\midrule
ZINC     & 0.4$\%$ & 0.919 &$1.3\%$ & 0.901 \\
\midrule
JT-VAE   & 11.3$\%$ & 0.846 &$7.6\%$ & 0.907 \\
ORGAN   & 0 & $-$ &$0.2\%$ & \bf{0.909} \\
GCPN   & $85.5\%$ & $0.392$ &$54.7\%$ & 0.855 \\
LIMO   & $10.4\%$ & \bf{0.914} &$-$& $-$ \\
\midrule
\textbf{SGDS}     & $\mathbf{86.0\%}$  & 0.874 &$\mathbf{62.2\%}$ & 0.858\\
\end{tabular}
}
\end{table}

\subsection{Ablation Studies}
SGDS outperforms previous methods by a significant margin, especially on binding affinity related experiments. Hence, we conduct ablations on the key components of our method on a challenging single-objective  ACAA1 maximization experiment. Since SGDS is optimized based on shifting the joint distribution rather than per-molecule based optimization method (such as \citep{eckmann2022limo}), we use the summarized statistics (i.e. the mean and standard deviation of $100$ lowest $\mathrm{K_d}$ from uniquely generated molecules) of last three shifted distributions as our metric to compare the key components rather than top-3 optimized $\mathrm{K_D}$ in \mysecref{sub:single_property_optimization}. Ablation studies are discussed as follows.

(1) \textit{Without EBM Prior}: for joint distribution, we replace the learnable EBM prior by a fixed $\mathcal{N}(0,I_d)$.  (2)\textit{Without Property Regression} $p_\gamma(y|z)$: we only learn the distribution of molecules as $p_\alpha(z)p_\beta(x|z)$. For each iteration of distribution shifting, we only use rank-and-select and update model parameters based on those molecules with high values. The molecule can be generated by first sampling $z\sim p_\alpha(z)$ and then $x\sim p_\beta(x|z)$. (3) \textit{Without Gradual Shifting}: rather than iterative distribution shifting in SGDS, we directly sample $z\sim p_\theta(z|y=y^\star)$, where $y^\star$ is set to be the minimal value we can get in \mysecref{sub:single_property_optimization}. (4) \textit{Without Rank-and-Select:} we skip the rank-and-select step in \cref{algo:sgds}.
(5) \textit{Without Warm Start:} when sampling $z\sim p_\theta(z|y)$ in current iteration, we replace the warm start algorithm in \cref{eq:warm_start} by 20-step langevin dynamics in \cref{eq:Langevin} with the same step size.

\begin{table}[H]
\caption{Ablation Studies. Report the mean and standard deviation of $100$ uniquely generated molecules with the lowest $\mathrm{K_D}$(in $10^{-9}$mol/L) from last three shifted iterations (i.e. the 28th, 29th, 30th iterations of total 30 iterations).}
\label{tab:abla}
\resizebox{\linewidth}{!}{
\begin{tabular}{lccc}
\toprule
\multirow{1}{*}{\bf Method} & 28th       & 29th       & 30th       \\
\midrule
SGDS        & $0.74\pm 0.04$            & $0.61\pm 0.03$          & $0.59\pm 0.03$\\         
Without EBM Prior   & $47.8\pm 26.9$            & $38.8\pm 21.1$          &$35.1\pm 20.2$    \\
Without Property Regression    & $140\pm 74.8$            & $114\pm 67.0$          & $103\pm 56.3$ \\
Without Gradual Shifting    & $211\pm 125$            & $166\pm 97.9$          &$137\pm 74.8$ \\
Without Rank-and-Select    & $9.71\pm 5.52$            & $5.75\pm 3.39$          & $3.91\pm 2.17$\\
Without Warm Start    & $6.27\pm 3.92$            & $3.27\pm 2.99$          & $2.37\pm 1.35$\\
\end{tabular}
}
\end{table}
The ablation studies are displayed in \cref{tab:abla}.  It
is clear that all the proposed components contribute significantly to the good performance of our
method. 

\subsection{Unconditional Generation}
\label{sec:a2}
We employ unconditional molecule generation tasks as a sanity check of the latent space EBM. The goal is to model the molecules in the training dataset and generate similar molecules. We evaluate the model based on validity (the percentage of molecules that satisfy the chemical valency rules), uniqueness (the percentage of unique molecules in all generated samples) and novelty (the percentage of generated molecules that are not in the training set) of generated molecules. Note that we are not concerned with optimization of molecular properties in this subsection.


Following previous work, we randomly sample $10$k molecules for ZINC and $30$k for MOSES, comparing the results based on the aforementioned metrics. Generation results are shown in Table~\ref{tab:gen_on_zinc} for ZINC and Table~\ref{tab:gen_on_moses} for MOSES.
In Table~\ref{tab:gen_on_zinc}, we present generation results for both SMILES and SELFIES. Despite lacking a validity constraint during generation, our model attains $95.5\%$ validity using SMILES, outperforming other SMILES-based methods and rivaling those with valency checks. This demonstrates our model's effective and implicit capture of valency rules. Furthermore, our model's samples exhibit perfect uniqueness and novelty.
\begin{table}[H]
\caption{Unconditional generation on ZINC. $^\star$ denotes valency check. \citep{jin2018junction,you2018graph,madhawa2019graphnvp,shi2020graphaf,luo2021graphdf,gomez2018automatic,kusner2017grammar} }
\label{tab:gen_on_zinc}
\resizebox{\linewidth}{!}{
\begin{tabular}{lcccc}
    \toprule
    \textbf{Model} & \textbf{Representation} & \textbf{Validity} & \textbf{Novelty} & \textbf{Uniqueness}\\
    \midrule
    JT-VAE   & Graph & 1.000$^\star$  & 1.000 & 1.000\\
    GCPN     & Graph & 1.000$^\star$  & 1.000 & 1.000\\
    GraphNVP  & Graph &  0.426 & 1.000 & 0.948 \\
    GraphAF       & Graph & 1.000$^\star$ & 1.000 & 0.991 \\
    GraphDF      & Graph    & {1.000}$^\star$   & 1.000 & 1.000\\
    \midrule
    ChemVAE     & \texttt{SMILES}    & 0.170 & 0.980 & 0.310\\
    GrammarVAE  & \texttt{SMILES}    & 0.310 & 1.000 & 0.108\\
    \textbf{Ours}  & \texttt{SMILES}    & 0.955  & {1.000} & {1.000}\\
    \textbf{Ours}  & \texttt{SELFIES}    & {1.000}  & {1.000} & {1.000}
\end{tabular}
}
\end{table}

\begin{table}[H]
\caption{Unconditional generation on MOSES. $^\star$ denotes valency check. Results obtained from \citep{polykovskiy2020molecular,eckmann2022limo}.}
\label{tab:gen_on_moses}
\resizebox{\linewidth}{!}{
\begin{tabular}{lclcc}
\toprule
\textbf{Model}                 & \textbf{Representation} & \textbf{Validity} & \textbf{Novelty} & \textbf{Uniqueness} \\ \midrule
JT-VAE & Graph                   & 1.000$^\star$     & 0.914            & 1.000               \\
GraphAF & Graph                   & 1.000$^\star$     & 1.000            & 0.991               \\
GraphDF & Graph                   & {1.000}$^\star$   & 1.000            & 1.000               \\
LIMO                           & \texttt{SELFIES}        & 1.000             & 1.000            & 0.976               \\
\textbf{Ours}                & \texttt{SELFIES}        & {1.000}           & {1.000}          & {1.000} 
\end{tabular}
}
\end{table}

\begin{figure}[H]
    \centering
    \includegraphics[width=0.45\textwidth]{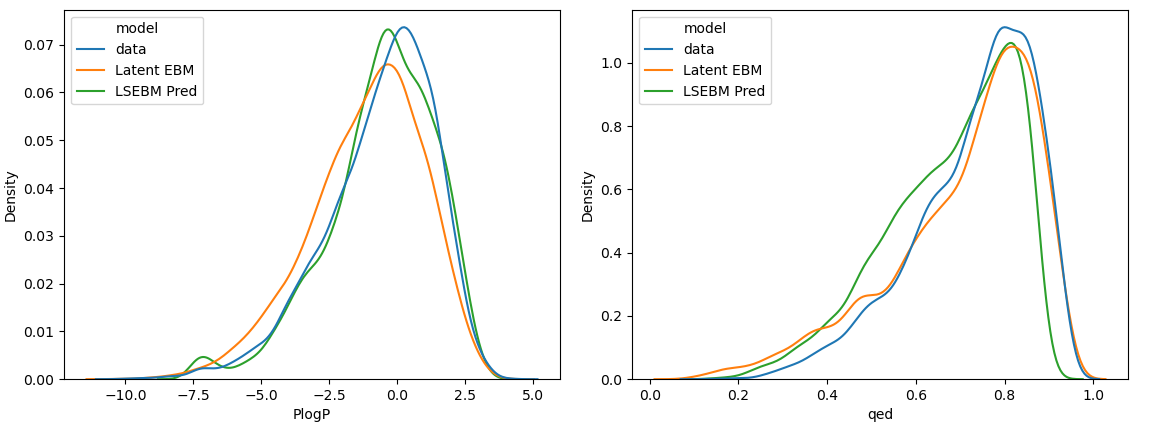}
    \caption{Property distributions of PlogP (left) and QED (right).}
    \label{fig:uncond}
\end{figure}

Then we randomly sample $10$k molecules from the learned latent space EBM and compute their PlogP and QED using RDKit. Their empirical densities are then compared not only with the molecule property densities from the test split but also with the predictions made by the regression model $p_\gamma(y|z)$. As shown in Figure~\ref{fig:uncond}, the property densities from both our learned model and predicted values from regression model align closely with those of the data, suggesting that our model effectively captures regularities in the data.

\begin{figure}[H]
\centering
    \includegraphics[width=0.48\textwidth]{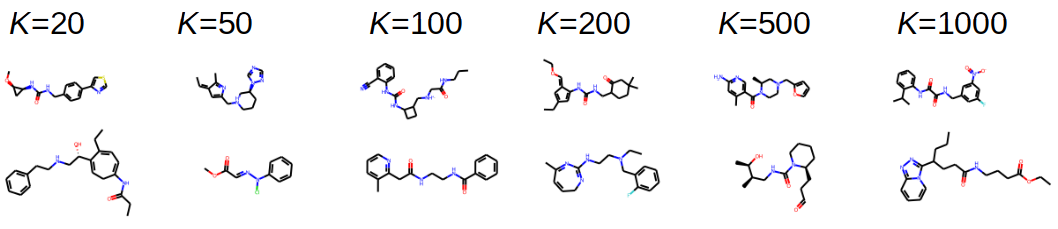}
    \caption{Two sequences of sampled molecules with different lengths of Markov chain.}
    \label{fig:langevin}
\end{figure}

In our experiments, we employ short-run MCMC \citep{nijkamp2019learning} with a Markov chain length of $K=20$ and step size $s=0.1$ for all tests. As shown in Figure~\ref{fig:langevin}, with increasing Markov chain length, the molecules evolve correspondingly, suggesting that the Markov chain is not trapped in local modes.
 
\section{Conclusion and discussion}
We propose a deep generative model for the joint distribution of molecules and their properties. It assumes an energy-based prior model in a low-dimensional continuous latent space, and the latent vector can generate the molecule and predict its property value.  
We then design a sampling with gradual distribution shifting method to shift the learned distribution to a region with high property values. Molecule design can then be achieved by conditional sampling. Our experiments demonstrate that our method outperforms previous SOTA methods on some tasks by significant margins. 


\begin{acknowledgements} 
Y. N. Wu was partially supported by NSF DMS-2015577 and a gift fund from Amazon.
\end{acknowledgements}

\bibliography{molecule_design}
\appendix
\onecolumn 

\section{Details about model and learning} 
\label{sec:a0}

Our model is of the form $p_\alpha(z)p_\beta(x|z) p_\gamma(y|z)$. The marginal distribution of $(x, y)$ is 
\begin{align*}
    p_\theta(x, y) = \int p_\theta(x, y, z) dz 
    = \int p_\alpha(z) p_\beta(x|z) p_\gamma(y|z) dz. 
\end{align*}
We use $p_\theta(x, y)$ to approximate the data distribution of $(x, y)$. 

For the data distribution of $(x, y)$, $y$ is a deterministic function of $x$. However, a machine learning method usually cannot learn the deterministic function exactly. Instead, we can only learn a probabilistic $p_\theta(y|x)$. Our model $p_\theta(x, y)$ seeks to approximate the data distribution $p(x, y)$ by maximum likelihood. A learnable and flexible prior model $p_\alpha(z)$ helps to make the approximation more accurate than a fixed prior model such as that in VAE. 

Let the training data be $\{(x_i, y_i), i = 1, ..., n\}$. The log-likelihood function is $L( \theta) = \sum_{i=1}^{n} \log p_\theta(x_i, y_i)$. The learning gradient is $L'( \theta) = \sum_{i=1}^{n} \nabla_\theta \log p_\theta(x_i, y_i)$. 
In the following, we provide details for calculating $\nabla_\theta \log p_\theta(x, y)$ for a single generic training example $(x, y)$ (where we drop the subscript $_i$ for notation simplicity).
\begin{align*} 
   \nabla_\theta  \log p_\theta(x, y) &= \frac{1}{p_\theta(x, y)} \nabla_\theta p_\theta(x, y)\\
    &= \frac{1}{p_\theta(x, y)} \int \nabla_\theta p_\theta(x, y, z) dz\\
    &= \frac{1}{p_\theta(x, y)} \int p_\theta(x, y, z) \nabla_\theta \log p_\theta(x, y, z) dz\\
    &=  \int \frac{p_\theta(x, y, z)}{p_\theta(x, y)} \nabla_\theta \log p_\theta(x, y, z) dz\\
    &=  \int p_\theta(z \mid x, y) \nabla_\theta \log p_\theta(x, y, z) dz\\
   & = \E_{p_\theta(z|x, y)} \left[ \nabla_\theta \log p_\theta(x, y, z) \right] \\
     &= \E_{p_\theta(z|x, y)} \left[ \nabla_\theta (\log p_\alpha(z) + \log p_\beta(x|z) + \log p_\gamma(y|z)) \right].
\end{align*}

For the prior model, 
\begin{align*}
    \nabla_\alpha \log p_\alpha(z) &= \nabla_\alpha f_\alpha(z) - \nabla_\alpha \log Z(\alpha) \\
    &= \nabla_\alpha f_\alpha(z) -\frac{1}{Z(\alpha)} \nabla_\alpha  Z(\alpha) \\
     &= \nabla_\alpha f_\alpha(z) -\frac{1}{Z(\alpha)}  \int  \nabla_\alpha \exp(f_\alpha(z)) p_0(z) dz \\
      &= \nabla_\alpha f_\alpha(z) -  \int  \nabla_\alpha f_\alpha(z) \frac{1}{Z(\alpha)} \exp(f_\alpha(z)) p_0(z) dz \\
    &=  \nabla_\alpha f_\alpha(z) - \E_{p_\alpha(z)}[ \nabla_\alpha f_\alpha(z)]. 
\end{align*}
Thus the learning gradient for $\alpha$ given an example $(x, y)$ is
\begin{align} 
\end{align}
The above equation has an empirical Bayes nature. $p_\theta(z|x, y)$ is based on the empirical observation $(x, y)$, while $p_\alpha$ is the prior model. 
For the generation model,
\begin{align} 
\begin{split}
  \delta_\beta(x, y) =  \nabla_\beta \log p_\theta(x, y)  = \E_{p_\theta(z|x, y)} [\nabla_\beta \log p_{\beta}(x|z)].
\end{split}
\end{align} 
Similarly, for the regression model, 
\begin{align} 
\begin{split}
  \delta_\gamma(x, y) =  \nabla_\gamma \log p_\theta(x, y)  = \E_{p_\theta(z|x, y)} [\nabla_\gamma \log p_{\gamma}(y|z)].
\end{split}
\end{align} 
Estimating expectations in the above equations requires Monte Carlo sampling of the prior model $p_\alpha(z)$ and the posterior distribution $p_\theta(z|x,y)$. If we can draw fair samples from the two distributions, and use these Monte Carlo samples to approximate the expectations, then the gradient ascent algorithm based on the Monte Carlo samples is the stochastic gradient ascent algorithm or the stochastic approximation algorithm of Robbins and Monro \citep{robbins1951stochastic}, who established the convergence of such an algorithm to a local maximum of the log-likelihood. 

For MCMC sampling using Langevin dynamics, the finite step or short-run Langevin dynamics may cause bias in Monte Carlo sampling. The bias was analyzed in \cite{pang2020learning}. The resulting algorithm is an approximate maximum likelihood learning algorithm.

\section{Training Time}
\label{sec:a3}

The training of joint distribution of molecule and its properties takes around 4 hours with $25$ iterations on a single Nvidia Titan XP GPU with batch size $2048$. For non-biological single-objective property optimization, it takes around $20$ minutes to do $30$ distribution shifting (SGDS) iterations. If we use SGDS without warm start, it takes around half an hour. For biological binding affinity maximization, the optimization time is mainly dependent on the number of queries of AutoDock-GPU. We do $30$ and $20$ SGDS iterations for the single-objective and multi-objective tasks, respectively, which cost $10$ hours and $8$ hours with warm start, and cost $14$ hours and $10$ hours without warm start. For biological property optimization tasks, we use two Nvidia Titan XP GPUs, one for running our code and the other for running AutoDock-GPU. We have added a table to compare with previous methods.

\begin{table}[H]
\small
\begin{center}
\begin{tabular}{lccc}
\textbf{Model}                 & Penalized-logP/QED  &Single binding affinity\\ \hline
JT-VAE &24  &$-$  \\
GCPN & 8   &6                \\
MolDQN & 24  &6                 \\
GraphDF & 8   &12               \\
Mars   & 12  &6                   \\
LIMO   & 1  & 1                       \\
SGDS without warm start& $4.5$ & $18$ \\
SGDS with warm start& $4.3$ & $14$ \\
\end{tabular}
\end{center}
\caption{\small Comparison of molecule generation time in (hrs). Results obtained from \citep{eckmann2022limo}.}
\label{tab:time}
\end{table}

Even if we use MCMC sampling-based methods, our training speed is affordable comparing to existing methods. That is because our designed latent space EBM is low-dimensional (i.e. dim$(z)$=100) and we use short-run MCMC (i.e. with fixed iteration steps $20$) in our experiments. The major bottleneck of the training speed is the time of querying the property compute engines.

\section{Generated Samples}
\subsection{Biological Property Optimization}

\begin{figure}[H]
    \centering
    \includegraphics[width=.7\textwidth]{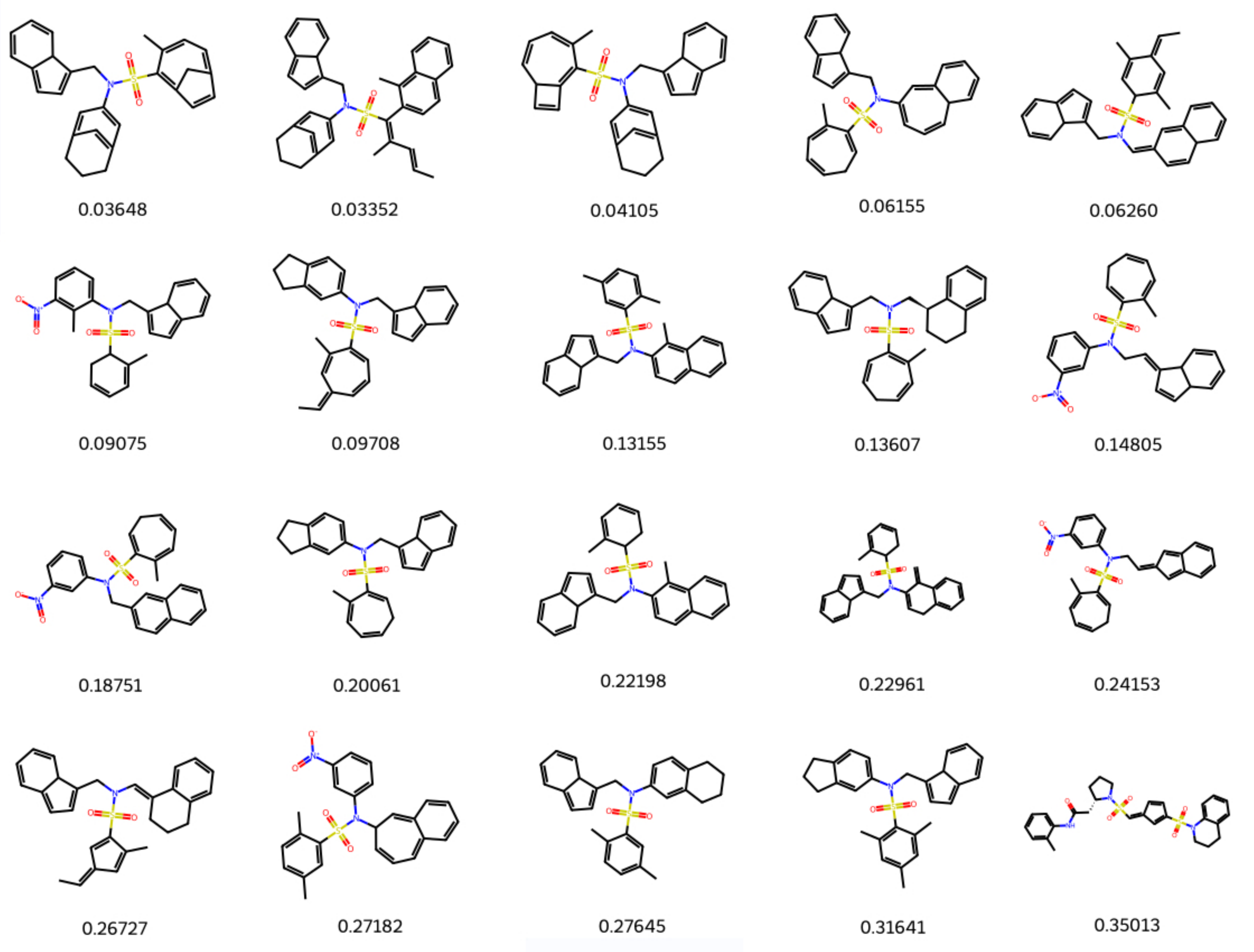}
    \caption{\small Generated molecules in {singe-objective} esr1 binding affinity maximization experiments with corresponding $\mathrm{K_D}(\downarrow)$ in nmol/L.}
    \label{fig:ba0}
\end{figure}

\begin{figure}[H]
    \centering
    \includegraphics[width=.8\textwidth]{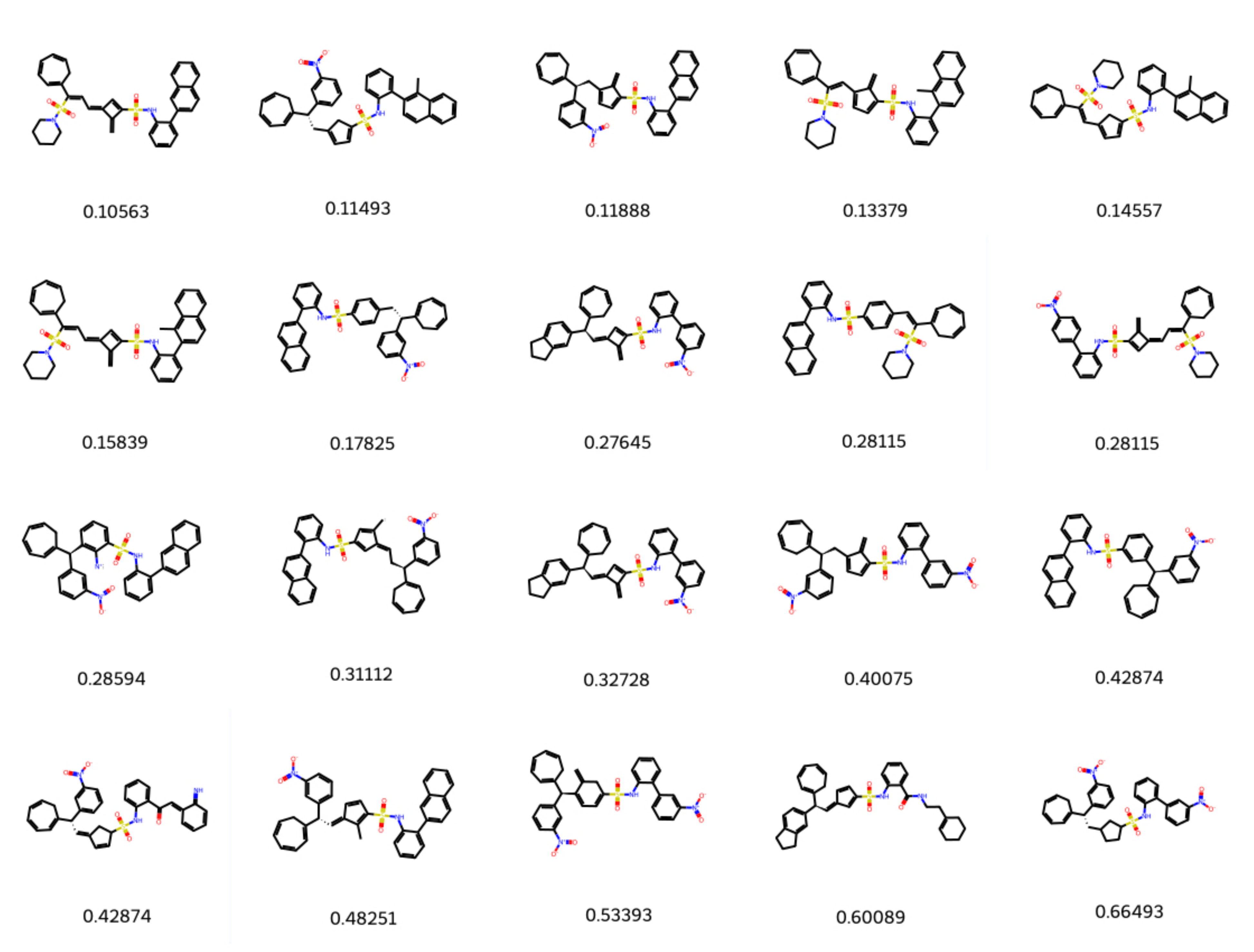}
    \caption{\small Generated molecules in {singe-objective} acaa1 binding affinity maximization experiments with corresponding $\mathrm{K_D}(\downarrow)$ in nmol/L.}
    \label{fig:ba1}
\end{figure}

Figure~\ref{fig:ba0} and Figure~\ref{fig:ba1} show generated molecules with high binding affinities towards ESR1 and ACAA1 respectively in single-objective property design experiments. 

\begin{figure}[H]
    \centering
    \includegraphics[width=.75\textwidth]{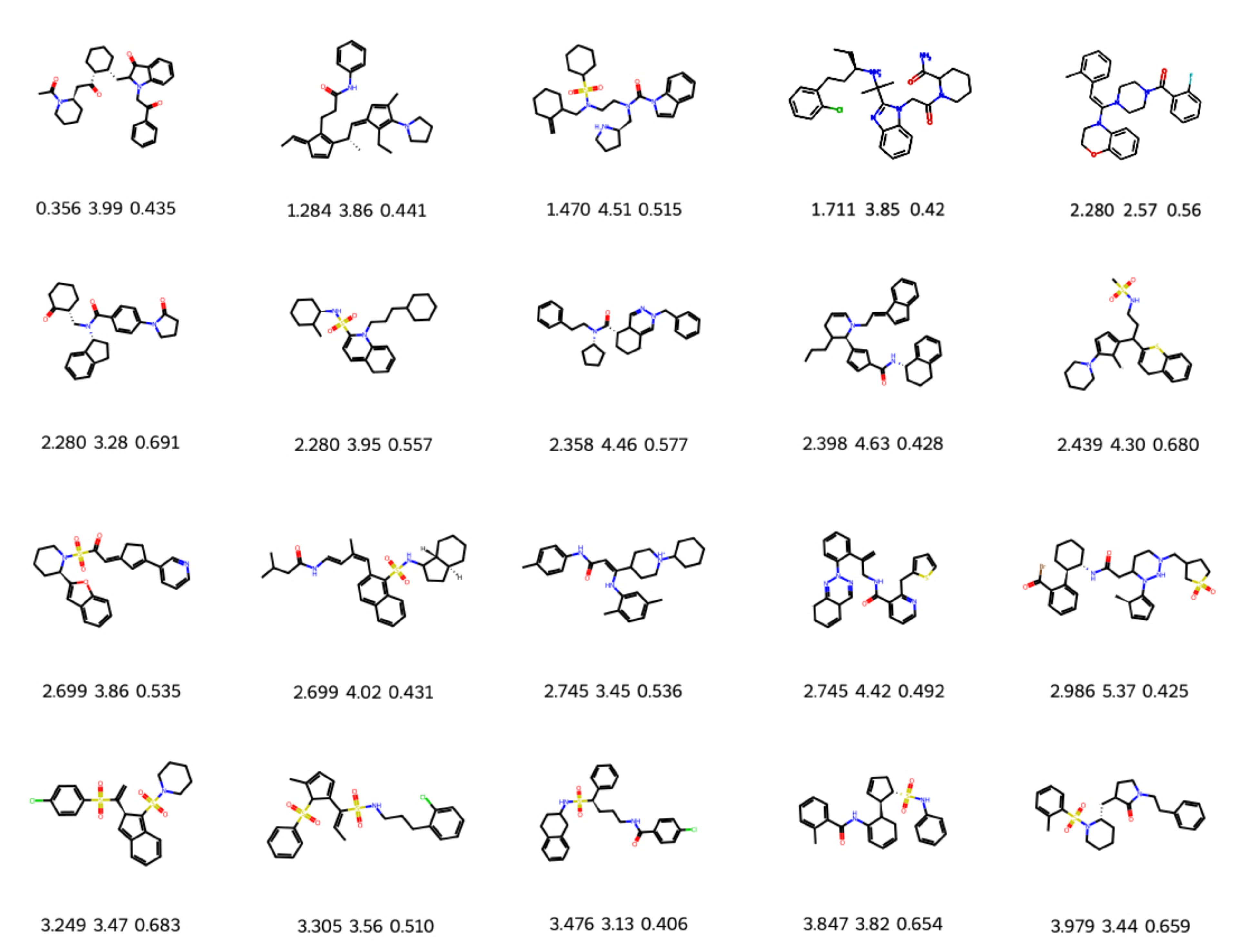}
    \caption{\small Generated molecules in {multi-objective} esr1 binding affinity maximization experiments with corresponding $\mathrm{K_D}(\downarrow)$ in nmol/L, SA$(\downarrow)$ and QED$(\uparrow)$ respectively.}
    \label{fig:mba0}
\end{figure}

\begin{figure}[H]
    \centering
    \includegraphics[width=.75\textwidth]{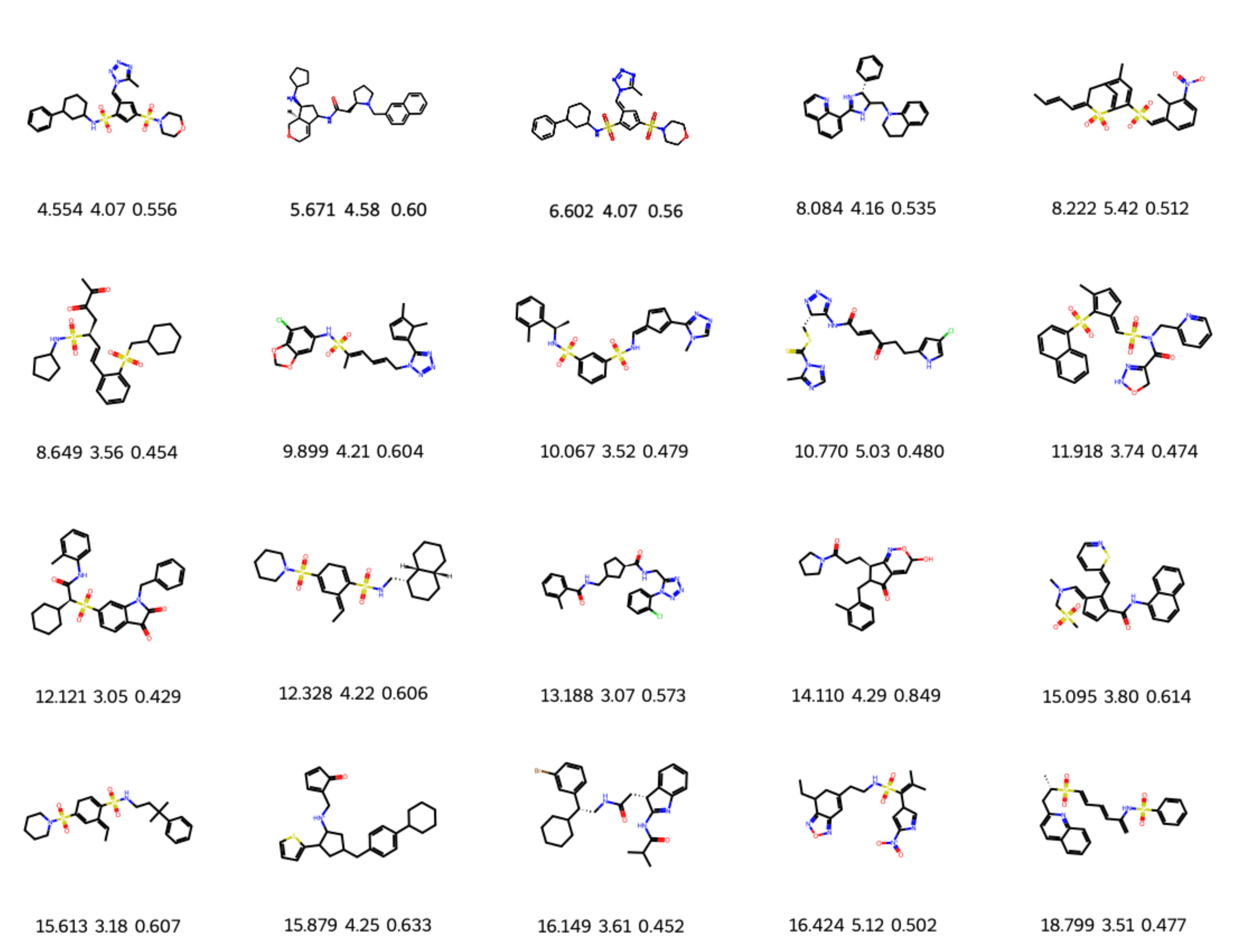}
    \caption{\small Generated molecules in {multi-objective} acaa1 binding affinity maximization experiments with corresponding $\mathrm{K_D}(\downarrow)$ in nmol/L, SA$(\downarrow)$ and QED$(\uparrow)$ respectively.}
    \label{fig:mba1}
\end{figure}

Figure~\ref{fig:mba0} and Figure~\ref{fig:mba1} show generated molecules with high binding affinities towards ESR1 and ACAA1 respectively in multi-objective property design.

Comparing to the previous state-of-the-art methods, SGDS is able to produce more high quality molecules than top-3 molecules because after gradual distribution shifting, the joint distribution locates at the area supported by molecules with high binding affinities.



Meanwhile, compared to previous generative model-based methods, we use Langevin dynamics to infer the posterior distribution $p(z|x,y_1,\dots, y_n)$ without bothering to design different encoders when facing different combination of properties.

\subsection{P-logP and QED Optimization}
\begin{figure}[H]
    \centering
    \includegraphics[width=.4\textwidth]{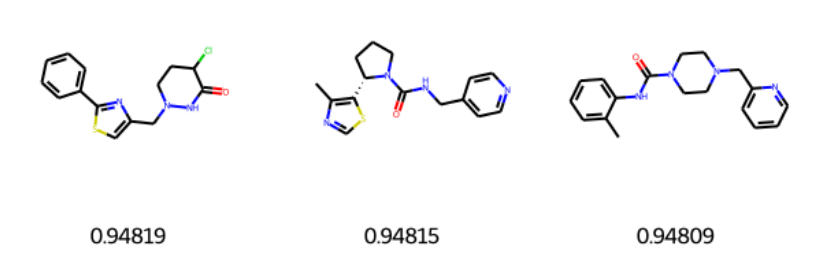}
    \caption{\small Top-3 molecules in single-objective QED maximization.}
    \label{fig:qed_single}
\end{figure}

\begin{figure}[H]
    \centering
    \includegraphics[width=0.6\textwidth]{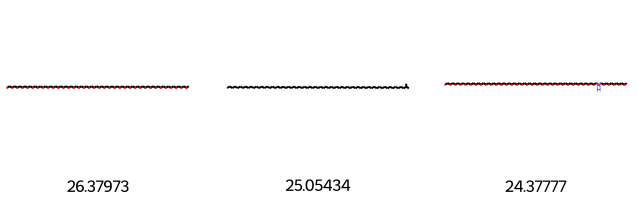}
    \caption{\small Top-3 molecules in single-objective p-logP maximization.}
    \label{fig:plogp_single}
\end{figure}

\newpage
\section{Illustration of Sampling with Gradual Distribution Shifting (SGDS)}
\label{sec:a1}

\begin{figure}[H]
    \centering
    \includegraphics[width=0.4\textwidth]{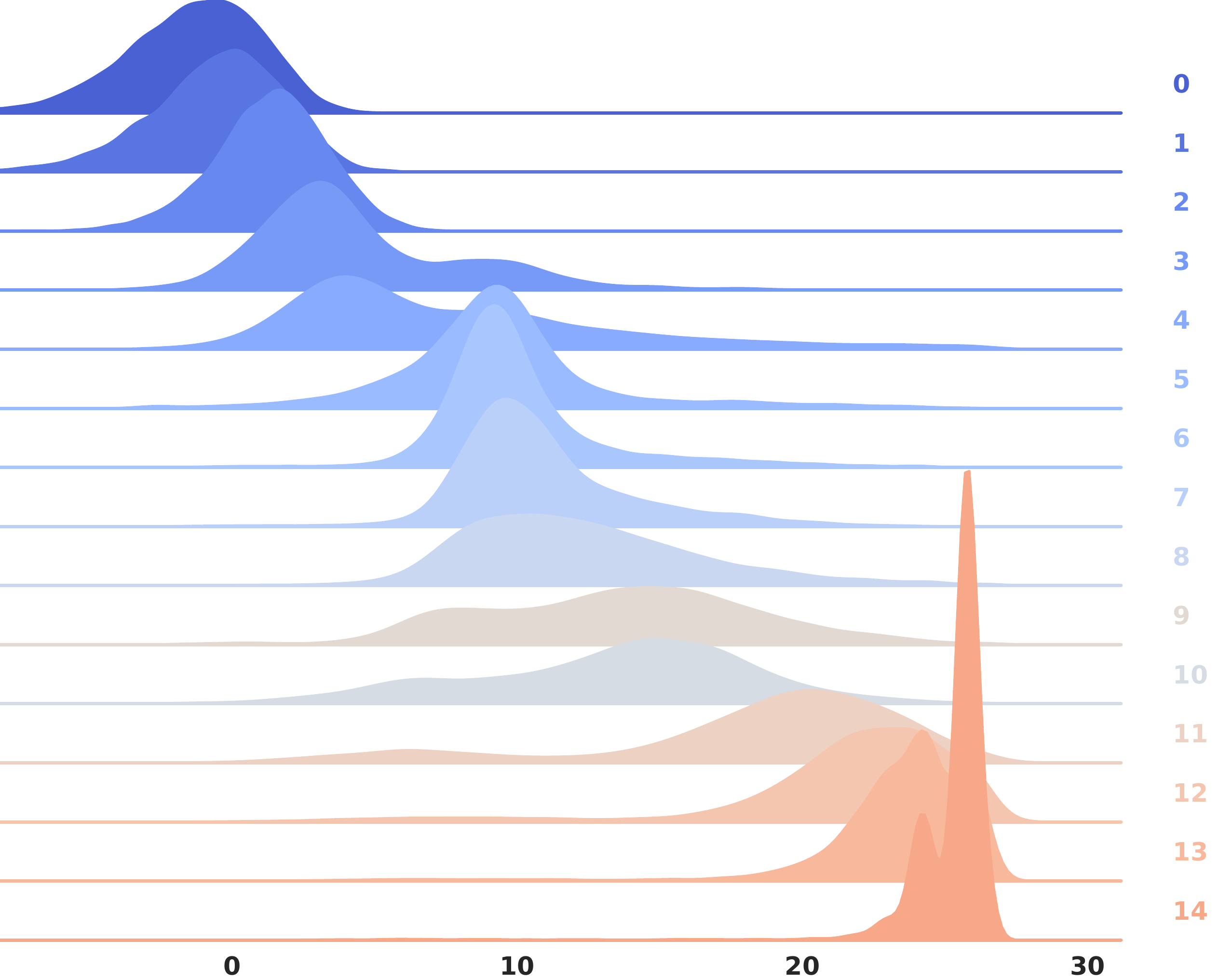}
    \caption{Illustration of SGDS in a single-objective penalized logP optimization experiment.}
    \label{fig:sgds}
\end{figure}

\begin{figure}[H]
    \centering
    \includegraphics[width=0.4\textwidth]{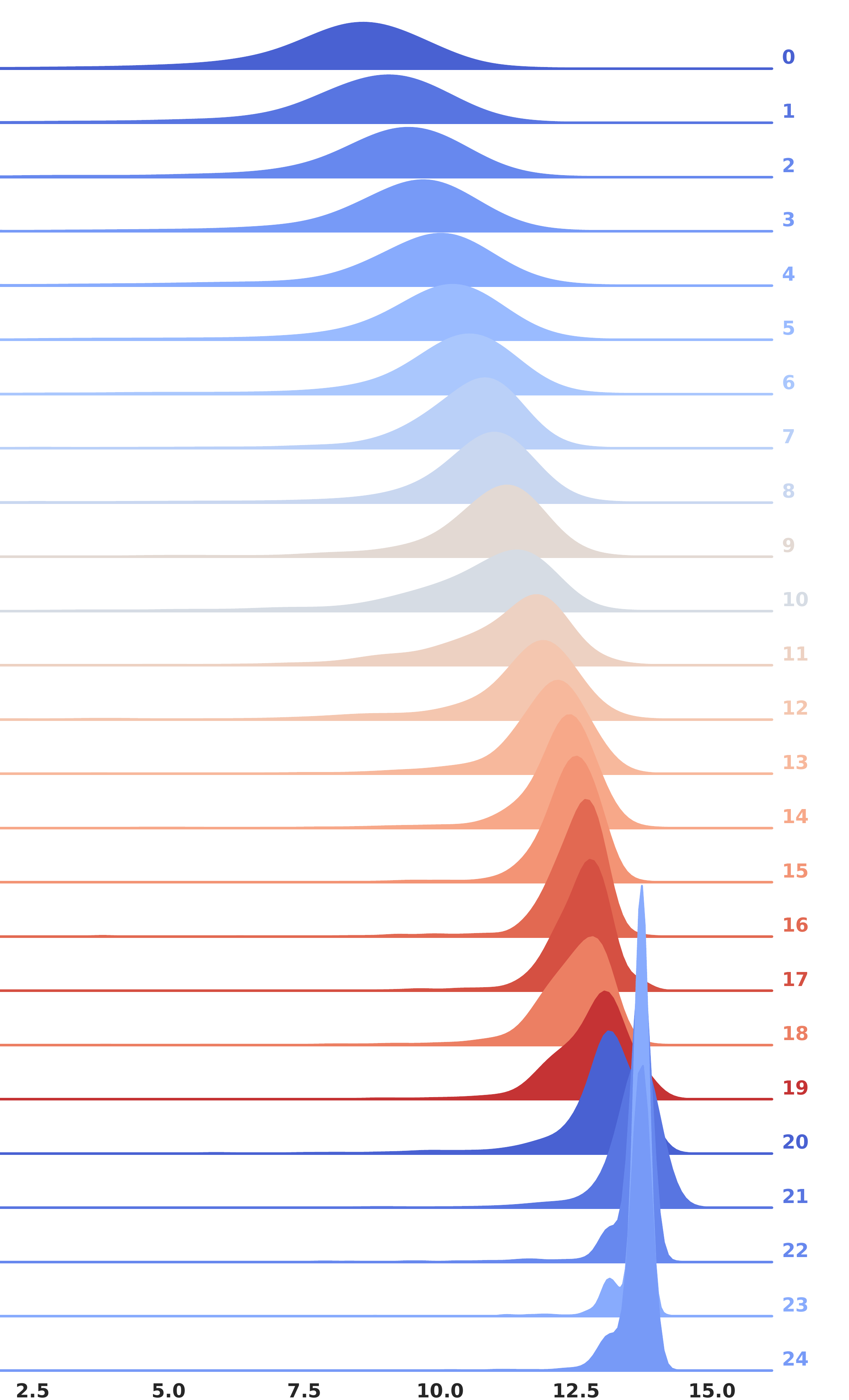}
    \caption{Illustration of SGDS in a single-objective esr1 optimization experiment.}
    \label{fig:sgds_esr}
\end{figure}

\begin{figure}[H]
    \centering
    \includegraphics[width=0.4\textwidth]{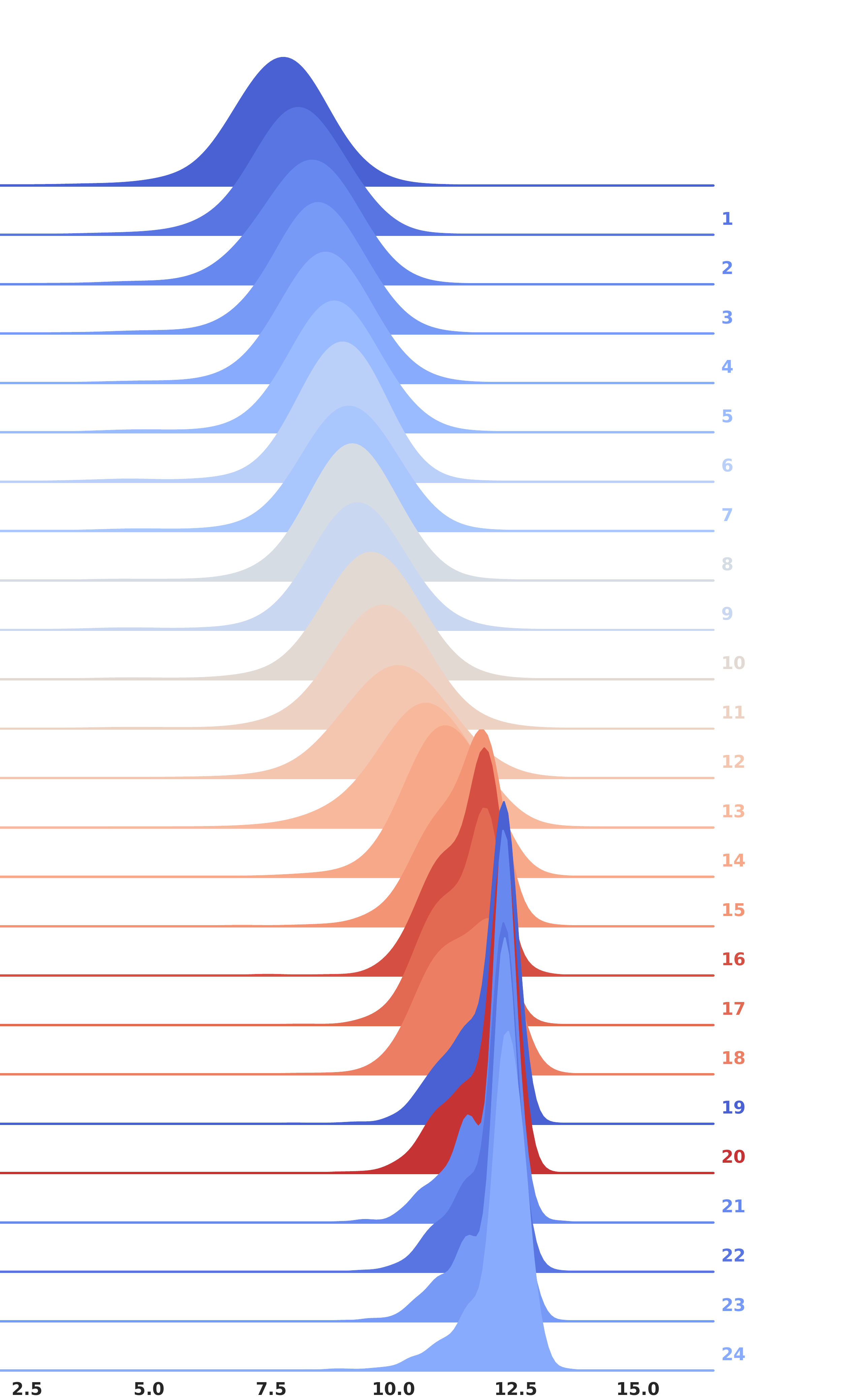}
    \caption{Illustration of SGDS in a single-objective acaa1 optimization experiment.}
    \label{fig:sgds_acaa}
\end{figure}

\cref{fig:sgds,fig:sgds_esr,fig:sgds_acaa} show property densities of sampled molecules of the distribution shifting process in single-objective penalized logP, esr1 and acaa1 optimization respectively. SGDS is implemented with warm start. We can see the model distribution is gradually shifting towards the region supported by molecules with high property values. To better visualize the shifting process, we plot the docking scores rather than $\mathrm{K_D}$. The increase in docking scores corresponds to the exponential decrease in $\mathrm{K_D}$.

\end{document}